\documentclass{article}

\usepackage[a4paper, portrait, margin=1in]{geometry}
\usepackage{graphicx} % Required for inserting images
\usepackage[colorinlistoftodos]{todonotes}
\usepackage{float}
\usepackage{multirow}
\usepackage[colorlinks, citecolor=cyan]{hyperref}

\usepackage[backend=biber,sorting=none,minbibnames=3, style=numeric-comp]{biblatex}
\usepackage{amsmath}
\AtBeginEnvironment{gather}{
\setlength{\abovedisplayskip}{0pt}
}
\usepackage{breqn}

\usepackage{lineno}
\usepackage[font=footnotesize]{caption}
\usepackage{subcaption}
\usepackage{titlesec}
\titlespacing\section{0pt}{12pt plus 4pt minus 2pt}{0pt plus 2pt minus 2pt}
\titlespacing\subsection{12pt}{12pt plus 4pt minus 2pt}{0pt plus 2pt minus 2pt}
\titlespacing\subsubsection{12pt}{12pt plus 4pt minus 2pt}{0pt plus 2pt minus 2pt}
\titleformat{\section}{\normalfont\fontsize{12}{15}\bfseries}{\thesection.}{1em}{}
\titleformat{\subsection}{\normalfont\fontsize{12}{15}\bfseries}{\thesubsection.}{1em}{}
\titleformat{\subsubsection}{\normalfont\fontsize{12}{15}\bfseries}{\thesubsubsection.}{1em}{}
\title{\textbf{\huge Reducing Artifacts in Grating Interferometry Using Multiple Harmonics and Phase Step Corrections}}

\usepackage{authblk}

\author[1]{Hunter C. Meyer}
\author[2]{Conner B. Dooley}
\author[1]{Victoria L. Fontenot}
\author[3]{Kyungmin Ham}
\author[4]{Leslie G. Butler}
\author[5]{Alexandra Noël}
\author[1,*]{Joyoni Dey}
\affil[1]{Department of Physics and Astronomy, Louisiana State University, Baton Rouge, LA, 70803}
\affil[2]{Department of Physics, Linfield University, McMinnville, OR, 97128}
\affil[3]{Center for Advanced Microstructures and Devices, Louisiana State University, Baton Rouge, LA, 70806}
\affil[4]{Department of Chemistry, Louisiana State University, Baton Rouge, LA, 70803}
\affil[5]{Department of Comparative Biomedical Sciences, Louisiana State University School of Veterinary Medicine, Baton Rouge, LA, 70803}
\affil[*]{Corresponding Author: deyj@lsu.edu}

\date{} % Need to do this or else the date is shown

\begin{document}

\maketitle
\vspace{-3em} % Reduce space between affiliations and abstract

\begin{abstract}

X-ray interferometry is an emerging imaging modality with a wide variety of potential clinical applications, including lung and breast imaging, as well as in non-destructive testing, such as additive manufacturing and porosimetry.  A grating interferometer uses a diffraction grating to produce a periodic interference pattern and measures how a patient or sample perturbs the pattern, producing three unique images that highlight X-ray absorption, refraction, and small angle scattering, known as the transmission, differential-phase, and dark-field images, respectively.  Image artifacts that are unique to X-ray interferometry are introduced when assuming the fringe pattern is perfectly sinusoidal and the phase steps are evenly spaced.  Inaccuracies in grating position, coupled with multi-harmonic fringes, lead to remnant oscillations and phase wraparound artifacts.  We have developed an image recovery algorithm that uses additional harmonics, direct relative phase fitting, and phase step corrections to prevent them.  The direct relative phase fitting removes the phase wraparound artifact.  Correcting the phase step positions and introducing the additional harmonic removes the grating remnant artifact present in the transmission, differential-phase, and dark-field images.  By modifying existing algorithms, the fit to the fringe pattern is greatly improved and artifacts are minimized, as we demonstrate with the imaging of several samples, including PMMA microspheres, ex vivo formalin fixed mouse lungs, and porous alumina.

\end{abstract}

\section{Introduction}
\label{sec:introduction}

X-ray interferometry is a developing imaging modality with a wide variety of potential clinical and industrial applications, including lung imaging \cite{bib:Gassert, bib:Bech, bib:Yaroshenko, bib:Velroyen}, breast imaging \cite{bib:Wang2014, bib:Tapfer, bib:Scherer, bib:Koehler}, pore size analysis \cite{bib:Revol, bib:MeyerDeySciRep2024}, and additive manufacturing quality assurance \cite{bib:Zhao, bib:Brooks}.  A grating interferometer works by producing a periodic fringe pattern by placing a diffraction grating between an X-ray source and detector.  Three unique images are produced by measuring perturbations in the fringe pattern \cite{bib:Momose2005, bib:Pfeiffer2009}.  X-ray absorption decreases the average value of the interference fringes, producing the transmission image, which is the same as a traditional X-ray.  Refraction produces a lateral shift in the pattern, producing the differential-phase image.  Small angle X-ray scatter reduces the fringe visibility, producing the dark-field image.  The amount of dark-field signal depends on the system's autocorrelation length (ACL) and microstructures of the sample being imaged \cite{bib:Yashiro2010, bib:Strobl2014, bib:Gkoumas2016}.

Grating interferometers commonly use a technique known as phase stepping to acquire images.  The grating is translated laterally and imaged multiple times to produce a phase stepping curve at each pixel.  Two phase stepping curves are acquired, one with no sample that serves as a reference curve and one with a sample in place, and the fringes are analyzed and compared on a per-pixel basis to simultaneously produce the three images.  The phase stepping curves are typically assumed to be sinusoidal, and the average value, visibility, and phase are extracted and compared to produce the three images.  Phase stepping requires the use of high precision electronic motors, since the phase steps are on the micron level and at least five phase steps are usually acquired.

A key problem with a grating interferometer is small inaccuracies in grating position and the fact that the fringes are not perfectly sinusoidal, leading to image artifacts that are unique to grating interferometry.  The artifacts appear as remnant fringes in the transmission, differential-phase, and dark-field images due to oscillations in the sinusoidal fit parameters since the fringes were non-sinusoidal to begin with and the true phase steps are not evenly spaced.  When the sample phase stepping curve is acquired and compared with the reference phase stepping curves, the oscillation in the fit parameters does not perfectly line up, leading to remnant fringes in the three images.  Additionally, the differential-phase image will have phase wraparound, leading to bright stripes in the images.

We propose the use of corrected phase step positions and multiple harmonics in the analysis of the phase stepping curves to decrease or completely remove the fit parameter oscillation which leads to remnant fringes in the transmission, differential-phase, and dark-field images.  We also propose the inclusion of the reference phase into the sample's phase stepping curve so that the relative phase is fit directly, preventing phase wraparound.  With the introduction of multiple harmonics and the phase wraparound correction, we redefine how our images are calculated in a simple and intuitive manner.  Previous work on Moiré artifacts resulting from phase stepping errors and dose fluctuations has been reported \cite{bib:Kaeppler, bib:Hauke, bib:Hashimoto, bib:Tao}. They require iterative approaches which could have convergence or bias issues, particularly in the differential-phase image if the average difference in the true and nominal phase step positions between the reference and sample phase stepping curves is not negligible.  The methods presented in this paper for correcting phase stepping errors are do not require iteration.  To the best of our knowledge, we are the first to analyze multiple harmonics and their influence on images produced with grating interferometers.

\section{Methods}
\label{sec:methods}

\begin{figure}
    \centering
    \includegraphics[keepaspectratio=true, width=0.6\textwidth]{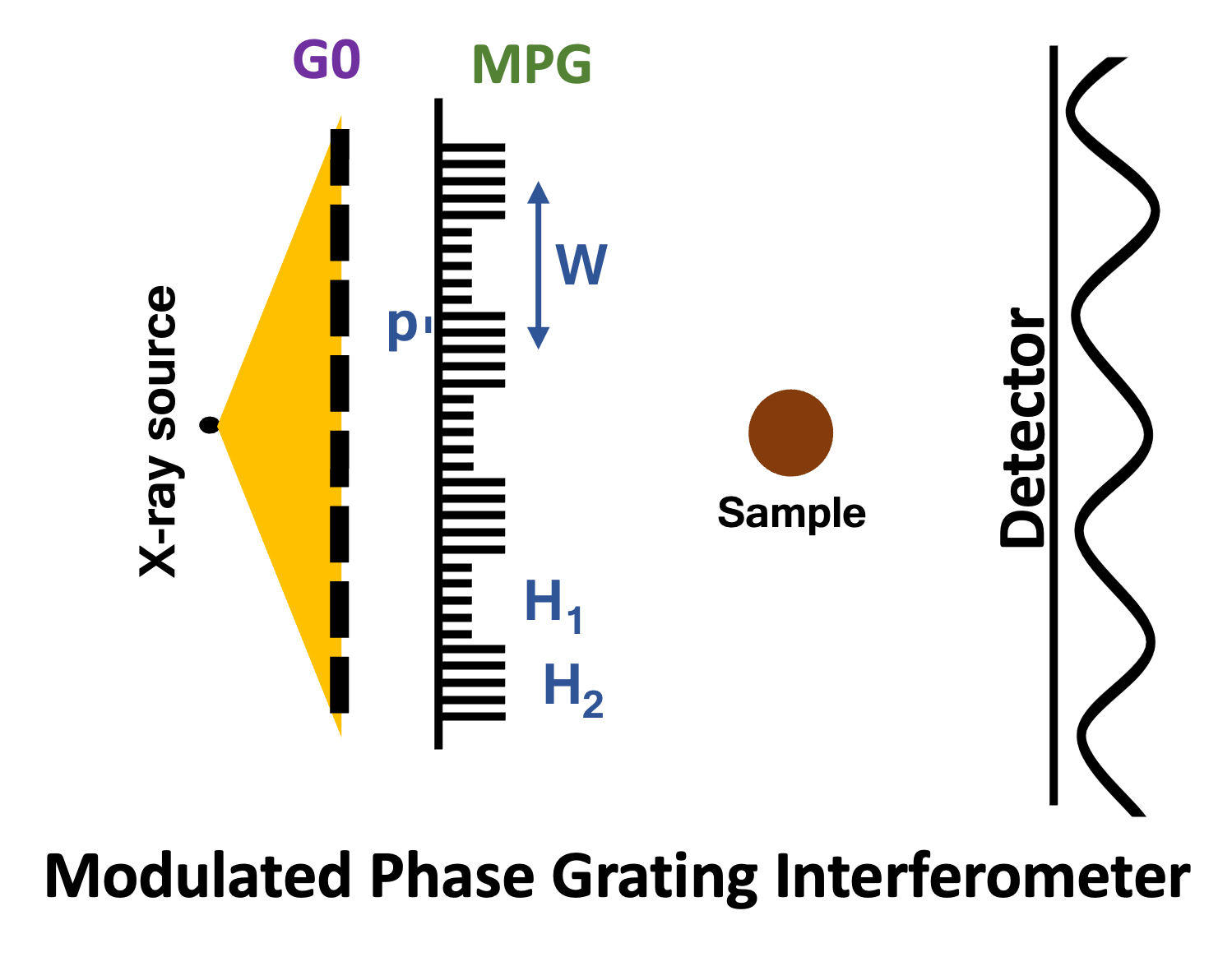}
    \caption{Schematic of the Modulated Phase Grating Interferometer, shown with a RectMPG.  The G0 grating is optional and not used in this study.}
    \label{fig:mpgi_schematic}
\end{figure}

The Modulated Phase Grating Interferometer (MPGI) is a recently developed X-ray system that is capable of interferometry without an analyzer grating \cite{bib:MPGPatent1, bib:MPGPatent2, bib:JXuHamDey, bib:HidrovoMeyerRSI, bib:MeyerDeySciRep2024}, shown in Figure \ref{fig:mpgi_schematic}.  The Modulated Phase Grating (MPG) is a diffraction grating that produces a periodic intensity pattern using either a microfocus source or standard X-ray source with a source grating, G0.  The MPG has grating bars separated by a pitch, $p$, and the bar heights follow an envelope function with period, $W$.  In the example shown, the envelope function is a RectMPG with phase-heights $(H_1, H_2)$.  The fringes produced by the MPG follow the period of the envelope function, $W$, and the smaller pitch, $p$, determines the coherence requirements to be in accordance with other X-ray interferometers.  Because $W$ is fairly large, the fringes produced by the MPG are directly resolvable for standard high resolution X-ray detectors, meaning no analyzer grating is required.

Transmission, differential-phase, and dark-field images are calculated by acquiring two phase-stepping curves: a reference curve and a sample curve.  A phase-stepping curve is a series of images generated by imaging the grating at several phase stepping positions, where the grating is laterally shifted over at least one period.  The phase stepping curves are analyzed on a pixel-by-pixel basis by fitting the measured intensity to a sinusoidal function shown in Equation \ref{eq:single_harmonic_fit}, where $x_g$ is the phase stepping position, $W$ is the period of the grating, and the fit parameters are $a_0$, $a_1$, and $\phi_1$, representing the average value, amplitude, and phase of the fringe pattern, respectively.  The subscript $1$ represents that only one harmonic was used, and the superscripts represent distinct equations for the reference, r, and sample, s.  The fringe visibility is the ratio of the amplitude and average value, $V = \frac{a_1}{a_0}$.  The phase stepping positions are typically assumed to be evenly spaced and go over at least one period.

\begin{equation}
    \label{eq:single_harmonic_fit}
    \hat{I}^{\: r,s}_1(x_g) = a_0^{r,s} + a_1^{r,s} \sin \left( \frac{2\pi x_g}{W} + \phi_1^{r,s} \right)
\end{equation}

The transmission, differential-phase, and dark-field images are found by comparing each pixel's fit parameters between the reference and sample acquisitions, shown in Equations \ref{eq:single_harmonic_transmission}, \ref{eq:single_harmonic_dpc}, and \ref{eq:single_harmonic_darkfield}.

\begin{gather}
    \label{eq:single_harmonic_transmission} 
    \text{Transmission} = \frac{a_0^s}{a_0^r} \\
    \label{eq:single_harmonic_dpc}
    \text{Differential-phase} = \phi_1^s - \phi_1^r \\
    \label{eq:single_harmonic_darkfield}
    \text{Dark-field} = \frac{V_1^s}{V_1^r}
\end{gather}

The parameters are easily fit by forming a linear least squares problem by separating $\phi_1^{r,s}$ from the $\frac{2\pi x_g}{W}$ using the angle addition formula and minimizing the sum squared error of each pixel between the predicted intensity, $\hat{I}^{\: r,s}_1(x_g)$, and the measured intensity, $I_m^{r,s}(x_g)$ \cite{bib:Marathe}.

There are multiple assumptions used in the analysis.  First, nominal values of $x_g$ are typically used.  That is to say, the phase steps are assumed to be evenly spaced.  However, inaccuracies in grating position resulting from motor positional accuracy or system vibrations lead to remnant oscillations in the fit parameters found in Equation \ref{eq:single_harmonic_single_image_fit}.  Additionally, we assumed that the fringe pattern was perfectly sinusoidal, with only a single harmonic, but diffraction is highly complex, with many harmonics contributing to the produced interference pattern \cite{bib:HidrovoMeyerRSI, bib:MeyerDeySciRep2024}.  Because we have a finite number of phase steps, the presence of multiple harmonics also leads to oscillations in the fitted visibility, even if the phase steps are truly evenly spaced.  Since the oscillations in the fit parameters are not perfectly aligned between the sample and reference fringe patterns, there are substantial remnant grating fringes in the transmission, differential-phase, and dark-field images.  Additionally, a phase wraparound artifact occurs when the fitted $\phi_1^{r,s}$ borders $\pm \pi$, since the phase gradients do not perfectly align between the reference and sample acquisitions.  The phase wraparound artifact presents itself as dark or brights streaks in the differential-phase image where the phase difference from Equation \ref{eq:single_harmonic_dpc} is equal to $\pm 2 \pi$.

\subsection{Phase Step Corrections}
\label{sec:methods_phase_step_corrections}

We account for inaccurate phase step positions by changing from using nominal values of $x_g$ to corrected values.  Using each phase step image, we estimate the phase step size between each step with the first reference phase step serving as the `zero' for both the reference and sample phase stepping curves.  This is important because there is no true zero for the phase steps, but the reference and sample curves must have the same `zero' to prevent a biased differential-phase image.

The phase step positions are corrected by fitting each phase step's 2-dimensional image to a sinusoidal pattern to find the phase difference relative to the previous step.  For the sample curve, the fringe pattern is of course altered by the sample, but for the purposes of fitting the entire image to a sine wave, the disturbances to the average value, visibility, and phase have a minimal impact on estimating the phase step size.  To prevent phase wraparound, the phase of the previous step is included in the fit of the next phase step.  As described earlier, for the first sample phase step, the \textit{phase of the previous step} is equal to the phase of the first reference step.  The fit equation is shown in Equation \ref{eq:single_harmonic_single_image_fit}.  The relative phase difference, $\alpha$, is found using linear least squares (with the reference phase included in the basis matrix, which will be described in more detail in Section \ref{sec:fitting_multiple_harmonics_and_relative_phase}), and the corrected phase step size is shown in Equation \ref{eq:corrected_phase_step_size}.  The corrected phase step size and reference phase of the next image is then updated, shown in Equations \ref{eq:update_phase_step} and \ref{eq:update_reference_phase}.

\begin{gather}
    \label{eq:single_harmonic_single_image_fit}
    \hat{I}_{2D}(x,y) = a_0 + a_1 \sin \left( \frac{2 \pi x}{W} + \phi^{\text{previous step}} + \alpha \right) \\
    \label{eq:corrected_phase_step_size}
    \Delta x_g = \frac{\alpha W}{2\pi} \\
    \label{eq:update_phase_step}
    x_g^{\text{next step}} = x_g^{\text{previous step}} + \Delta x_g\\
    \label{eq:update_reference_phase}
    \phi^{\text{next step}} = \phi^{\text{previous step}} + \alpha
\end{gather}

\subsection{Fitting Multiple Harmonics and Relative Phase}
\label{sec:fitting_multiple_harmonics_and_relative_phase}

In addition to correcting the phase step positions to reduce the grating remnant artifacts, further improvement is made by considering an additional harmonic.  We also fit the phase difference, $\Delta \phi_{1,2}$, directly by introducing the reference phase to the fit of the sample phase stepping curve, removing phase wraparound in the differential-phase images.  The new fit equations are shown in Equations \ref{eq:multiple_harmonics_fit_reference} and \ref{eq:multiple_harmonics_fit_sample}, where $A_{1,2}^{r,s}$ and $B_{1,2}^{r,s}$ are created using the angle addition formula and are combinations of the fringe amplitude and phase of each harmonic, $a_{1,2}^{r,s}$ and $\phi_{1,2}^{r}$ or $\Delta \phi_{1,2}$.

\begin{gather}
    \begin{split}
        \label{eq:multiple_harmonics_fit_reference}
        \hat{I}_2^r(x_g) & = a_0^r + a_1^r \sin \left( \frac{2 \pi x_g}{W} + \phi_1^r \right) + a_2^r \sin \left( \frac{4 \pi x_g}{W} + \phi_2^r \right) \\
        %& = a_0^r + a_1^r \cos \left( \phi_1^r \right) \sin \left( \frac{2 \pi x_g}{W} \right) + a_1^r \sin \left( \phi_1^r \right) \cos \left( \frac{2 \pi x_g}{W} \right) + a_2^r \cos \left( \phi_2^r \right) \sin \left( \frac{4 \pi x_g}{W} \right) + a_2^r \sin \left( \phi_2^r \right) \cos \left( \frac{4 \pi x_g}{W} \right) \\
        & = a_0^r + A_1^r \sin \left( \frac{2 \pi x_g}{W} \right) + B_1^r \cos \left( \frac{2 \pi x_g}{W} \right) + A_2^r \sin \left( \frac{4 \pi x_g}{W} \right) + B_2^r \cos \left( \frac{4 \pi x_g}{W} \right)
    \end{split} \\
    \begin{split}
        \label{eq:multiple_harmonics_fit_sample}
        \hat{I}_2^s(x_g) &= a_0^s + a_1^s \sin \left( \frac{2 \pi x_g}{W} + \phi_1^r + \Delta \phi _1 \right) + a_2^s \sin \left( \frac{4 \pi x_g}{W} + \phi_2^r + \Delta \phi_2 \right) \\
        %& = a_0^s + a_1^s \cos \left( \delta_1 \right) \sin \left( \frac{2 \pi x_g}{W} + \phi_1^r \right) + a_1^s \sin \left( \delta_1 \right) \cos \left( \frac{2 \pi x_g}{W} + \phi_1^r \right) + a_2^s \cos \left( \delta_2 \right) \sin \left( \frac{4 \pi x_g}{W} + \phi_2^r \right) + a_2^s \sin \left( \delta_2 \right) \cos \left( \frac{4 \pi x_g}{W} + \phi_2^r \right) \\
        &= a_0^s + A_1^s \sin \left( \frac{2 \pi x_g}{W} + \phi_1^r \right) + B_1^s \cos \left( \frac{2 \pi x_g}{W} + \phi_1^r \right) + A_2^s \sin \left( \frac{4 \pi x_g}{W} + \phi_2^r \right) + B_2^s \cos \left( \frac{4 \pi x_g}{W} + \phi_2^r \right)
    \end{split}
\end{gather}

The basis matrices used in linear least squares are shown in Equations \ref{eq:multiple_harmonics_basis_matrix_reference} and \ref{eq:multiple_harmonics_basis_matrix_sample}, with the second harmonic sine and cosines and the reference phase added for the sample, extending Marathe et al \cite{bib:Marathe}. Linear least squares is then performed, as shown in Equation \ref{eq:multiple_harmonics_linear_least_squares}.

\begin{gather}
    \label{eq:multiple_harmonics_basis_matrix_reference}
    M_2^r = \left[1, \cos \left( \frac{2 \pi x_g}{W} \right), \sin \left( \frac{2 \pi x_g}{W} \right), \cos \left( \frac{4 \pi x_g}{W} \right), \sin \left( \frac{4 \pi x_g}{W} \right) \right] \\
    \label{eq:multiple_harmonics_basis_matrix_sample}
    M_2^s = \left[ 1, \cos \left( \frac{2 \pi x_g}{W} + \phi_1^r \right), \sin \left( \frac{2 \pi x_g}{W} + \phi_1^r \right), \cos \left( \frac{4 \pi x_g}{W} + \phi_2^r \right), \sin \left( \frac{4 \pi x_g}{W} + \phi_2^r \right) \right] \\
    \label{eq:multiple_harmonics_linear_least_squares}
    \begin{bmatrix}
        a_0^{r} \\
        A_1^{r} \\
        B_1^{r} \\
        A_2^{r} \\
        B_2^{r}
    \end{bmatrix} = ({M_2^r}^T M_2^r)^{-1} {M_2^r}^T \: I_m^{r}(x_g)
    \: \text{, and }
    \begin{bmatrix}
        a_0^{s} \\
        A_1^{s} \\
        B_1^{s} \\
        A_2^{s} \\
        B_2^{s}
    \end{bmatrix} = ({M_2^s}^T M_2^s)^{-1} {M_2^s}^T \: I_m^{s}(x_g)
\end{gather}

The amplitude and phase of each harmonic are then extracted in a similar manner as before, except for the sample phase stepping curve, it's the relative phase differences as opposed $\phi_1^s$ or $\phi_2^s$, as shown in Equations \ref{eq:multiple_harmonics_amplitude}--\ref{eq:multiple_harmonics_phase_sample}.

\begin{gather}
    \label{eq:multiple_harmonics_amplitude}
    a_1^{r,s} = \sqrt{\left(A_1^{r,s}\right)^2 + \left(B_1^{r,s}\right)^2}
    , \:
    a_2^{r,s} = \sqrt{\left(A_2^{r,s}\right)^2 + \left(B_2^{r,s}\right)^2}
    \\
    \label{eq:multiple_harmonics_phase_reference}
    \phi_1^r = \tan^{-1} \left( \frac{B_1^{r}}{A_1^{r}} \right)
    , \:
    \phi_2^r = \tan^{-1} \left( \frac{B_2^{r}}{A_2^{r}} \right)
    \\
    \label{eq:multiple_harmonics_phase_sample}
    \Delta \phi_1 = \tan^{-1} \left( \frac{B_1^{s}}{A_1^{s}} \right)
    , \:
    \Delta \phi_2 = \tan^{-1} \left( \frac{B_2^{s}}{A_2^{s}} \right)
\end{gather}

With the introduction of multiple harmonics and the relative phase, the definition of the images must change.  The transmission image is exactly the same as before, just the ratio of the sample and reference $a_0$, as shown in Equation \ref{eq:multi_harmonic_transmission}.  The differential-phase image is now just the relative phase difference of the first harmonic, $\Delta \phi_1$, shown in Equation \ref{eq:multi_harmonic_dpc}.  We are currently ignoring $\Delta \phi_2$ for the differential-phase image.  For the dark-field image, the visibility can no longer be analytically computed.  Instead, the visibility must be found numerically.  Using the reference and sample fit parameters, the intensity is calculated at a high resolution (significantly higher than the step resolution) using Equations \ref{eq:multiple_harmonics_fit_reference} and \ref{eq:multiple_harmonics_fit_sample}, and the maximum and minimum intensity are found.  Then, the fringe visibility is numerically calculated, as shown in Equation \ref{eq:numerical_visibility}.  The dark-field image is then just the ratio of the sample and reference visibility, as shown in Equation \ref{eq:multi_harmonic_darkfield}.

\begin{gather}
    \label{eq:multi_harmonic_transmission} 
    \text{Transmission} = \frac{a_0^s}{a_0^r} \\
    \label{eq:multi_harmonic_dpc}
    \text{Differential-phase} = \Delta \phi_1 \\
    \label{eq:multi_harmonic_darkfield}
    \text{Dark-field} = \frac{V_{multi}^s}{V_{multi}^r} \\
    \label{eq:numerical_visibility}
    V_{multi} = \frac{I_{max} - I_{min}}{I_{max} + I_{min}}
\end{gather}

\subsection{Imaging Experiments}
\label{sec:methods_imaging}

To evaluate the proposed phase step corrections and algorithm changes, we have imaged multiple samples with a laboratory Modulated Phase Grating Interferometer.  A RectMPG is used, with design heights of $(\pi/2, \pi/8)$ at $25 \: keV$, a pitch of $p = 1 \: \mu m$, and envelope period of $W = 120 \: \mu m$.  For more information, see MPG8 from Meyer et al \cite{bib:MeyerDeySciRep2024}.  A source-to-detector distance of $110 \: cm$ is used, with a source-to-grating distance of $20 \: cm$.  A Hamamatsu L9181-02 microfocus X-ray source was used, so no G0 was necessary.  Images were acquired at $45 \: kVp$ and $55 \: \mu A$, under the small focal spot mode $(5 - 8 \: \mu m)$.  A Dexela 1512 X-ray detector was used with $75 \: \mu m \times 75 \: \mu m$ pixels, and phase stepping was performed with $12 \: \mu m$ phase steps, 20 second exposures, and 10 phase steps.

The proposed phase step corrections and algorithm changes were applied to three samples, including PMMA microspheres, an ex vivo formalin fixed mouse lung, and a porous alumina compound.  The source-to-grating distance was different for each sample, with $40 \: cm$ for the alumina and $15 \: cm$ for the PMMA and mouse lung.  PMMA microspheres have been shown to be an appropriate lung tissue analogue for dark-field imaging.  While $200 \: \mu m$ microspheres are an ideal lung tissue analogue \cite{bib:Spindler2023}, our microspheres were imaged with an autocorrelation length of $ACL = 60 \: nm$, so we used PMMA microspheres with a significantly smaller diameter of $1 \: \mu m$.  The porous alumina was a compound of SAS-90, a commercially available gamma alumina that's commonly used as a catalyst support.  The porous alumina also serves as a lung tissue analogue for dark-field imaging \cite{bib:MeyerDeySciRep2024}.  In addition to using lung tissue analogues, we used a mouse lung to assess the proposed phase step corrections and algorithm changes. We inflated and pressure-fixed (25 cm) a mouse lung with buffered formalin (10\%). The lungs were then used ex vivo for imaging.

\section{Results}
\label{sec:results}

\begin{figure}
    \begin{subfigure}[t]{0.49\textwidth}
        \centering
        \includegraphics[keepaspectratio=true, width=\textwidth]{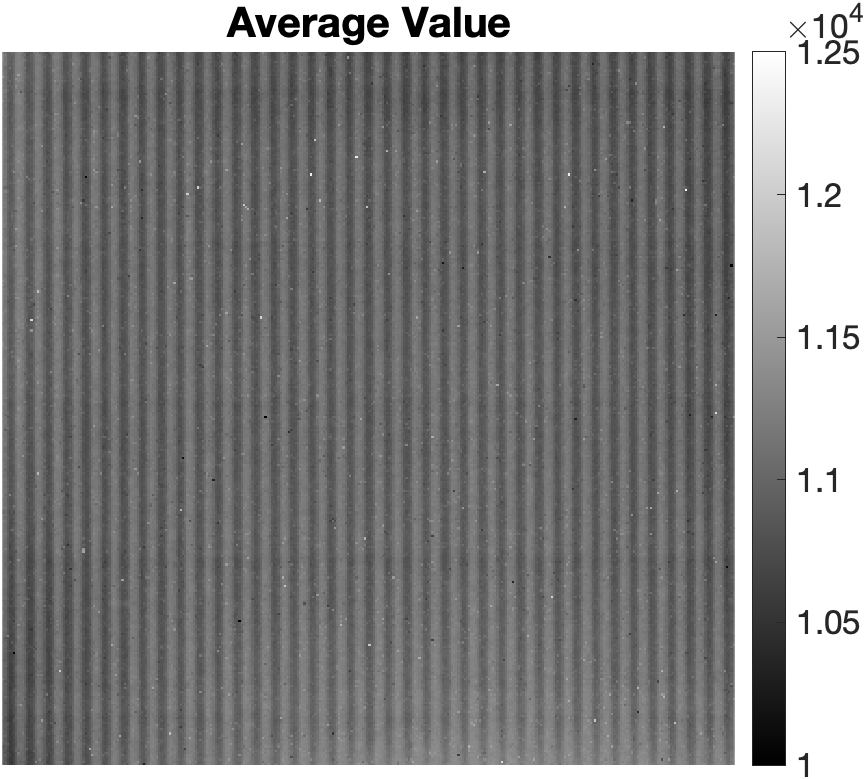}
        \subcaption{Nominal Steps}
        \label{fig:PMMA/a0_1st_only_nominal_s}
    \end{subfigure}
    \hfill
    \begin{subfigure}[t]{0.49\textwidth}
        \centering
        \includegraphics[keepaspectratio=true, width=\textwidth]{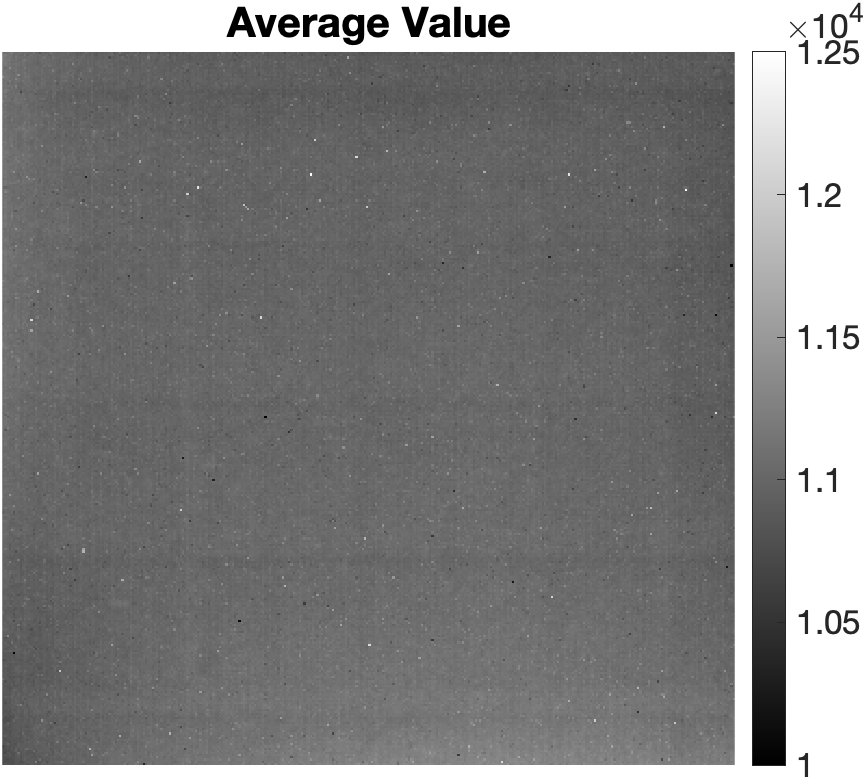}
        \subcaption{Corrected Steps}
        \label{fig:PMMA/a0_1st_only_corrected_s}
    \end{subfigure}
    \caption{Average value of the reference phase stepping curve for each pixel, $a^r_0$, calculated using (a) nominal phase steps and (b) corrected phase steps}
    \label{fig:PMMA/a0_comparison}
\end{figure}

\begin{figure}
    \begin{subfigure}[t]{0.49\textwidth}
        \centering
        \includegraphics[keepaspectratio=true, width=\textwidth]{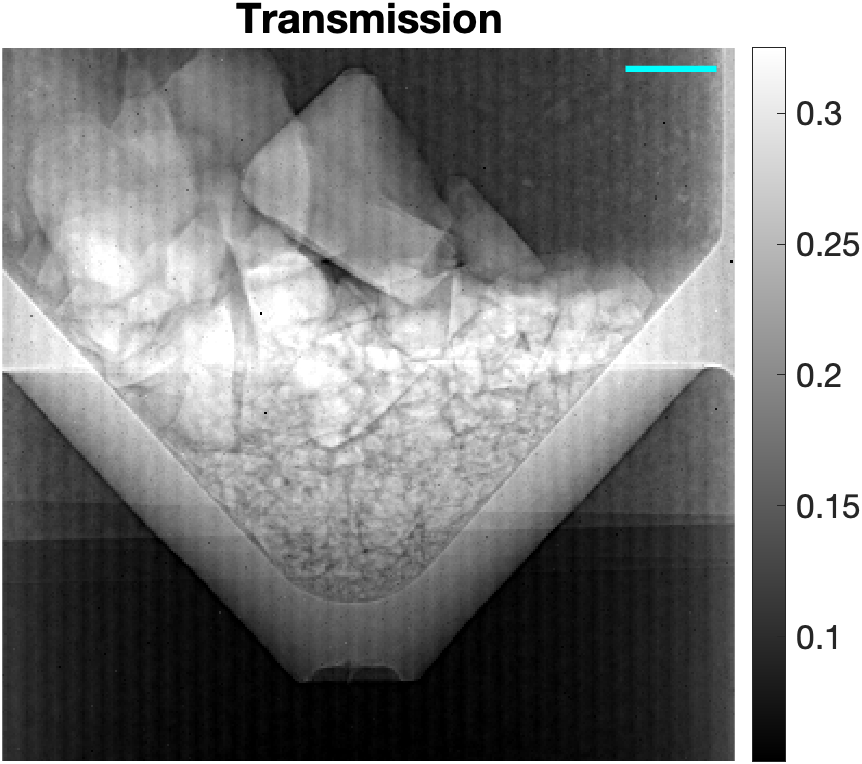}
        \subcaption{Nominal Steps}
        \label{fig:PMMA/transmission_nominal_s}
    \end{subfigure}
    \hfill
    \begin{subfigure}[t]{0.49\textwidth}
        \centering
        \includegraphics[keepaspectratio=true, width=\textwidth]{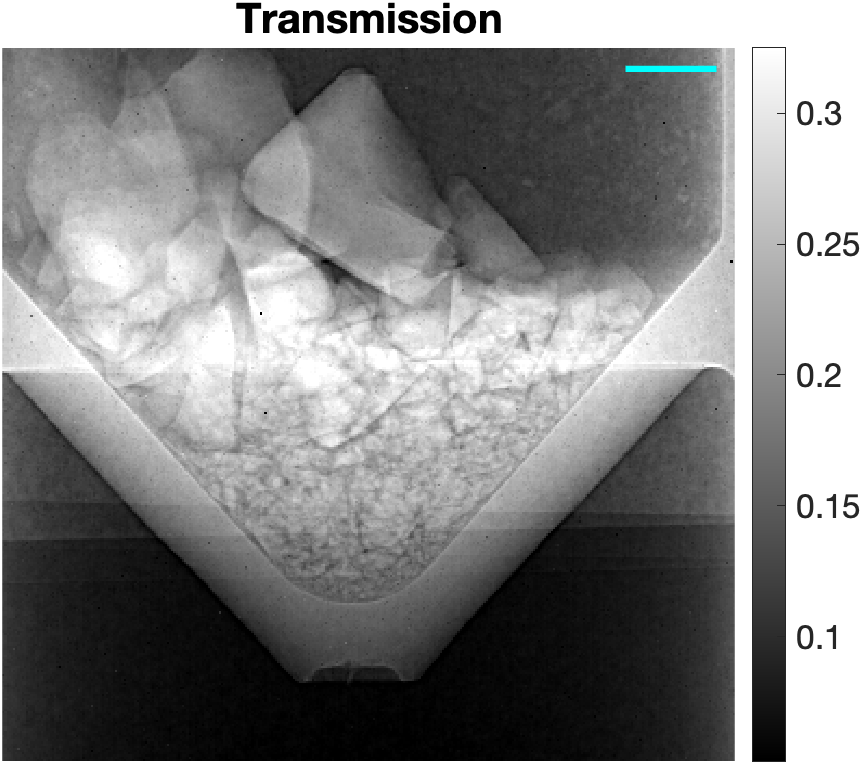}
        \subcaption{Corrected Steps}
        \label{fig:PMMA/transmission_corrected_s}
    \end{subfigure}
    \caption{Transmission Image shown on a -log scale calculated using (a) nominal phase steps and (b) corrected phase steps.  A $1 \: mm$ scalebar is shown.}
    \label{fig:PMMA/transmission_comparison}
\end{figure}

The effect of the phase step corrections, phase wraparound correction, and multiple harmonics on the transmission, differential-phase, and dark-field images are compared for the three samples.  The average value of each pixel's reference phase stepping curve was calculated using the nominal and corrected phase steps, shown in Figure \ref{fig:PMMA/a0_comparison}, with a noticeable reduction in remnant grating fringes.  The inclusion of the reference phase for phase wraparound correction and multiple harmonics have no effect on the average value, so are not shown.  With the reduction of the remnant fringes in the average value shown in Figure \ref{fig:PMMA/a0_comparison}, the effect this has on the transmission image of the PMMA microspheres is shown in Figure \ref{fig:PMMA/transmission_comparison}.  Again, a significant reduction in the remnant fringes is seen.  

\begin{figure}
    \begin{subfigure}[t]{0.49\textwidth}
        \centering
        \includegraphics[keepaspectratio=true, width=\textwidth]{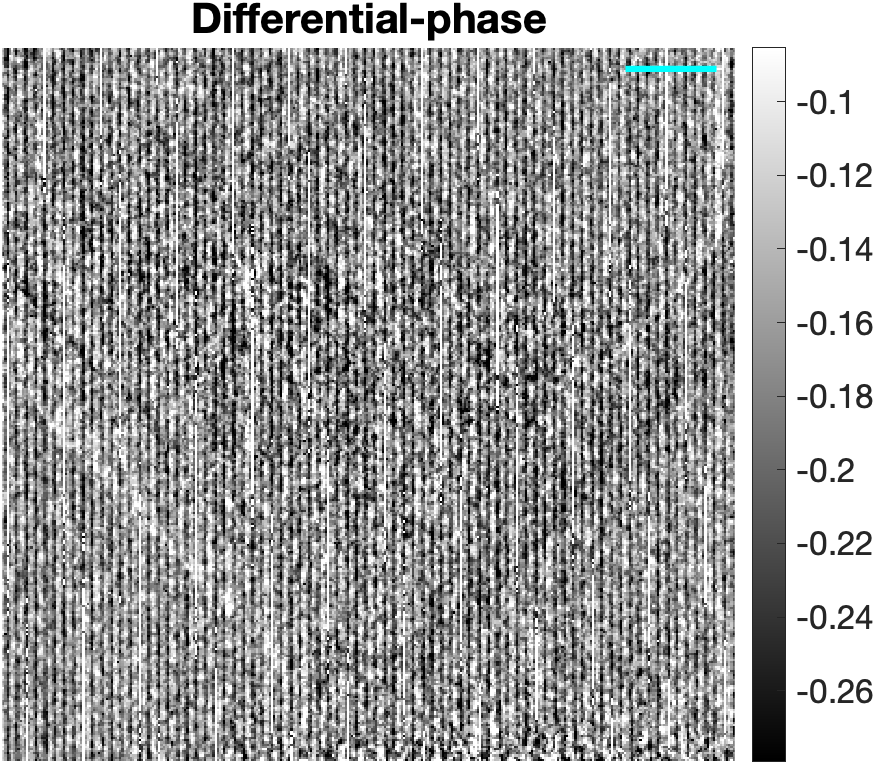}
        \subcaption{1st Harmonic, Nominal Steps, w/o Ref. Phase}
        \label{fig:PMMA/dpc_1st_only_nominal_s_no_ref_phase}
    \end{subfigure}
    \hfill
    \begin{subfigure}[t]{0.49\textwidth}
        \centering
        \includegraphics[keepaspectratio=true, width=\textwidth]{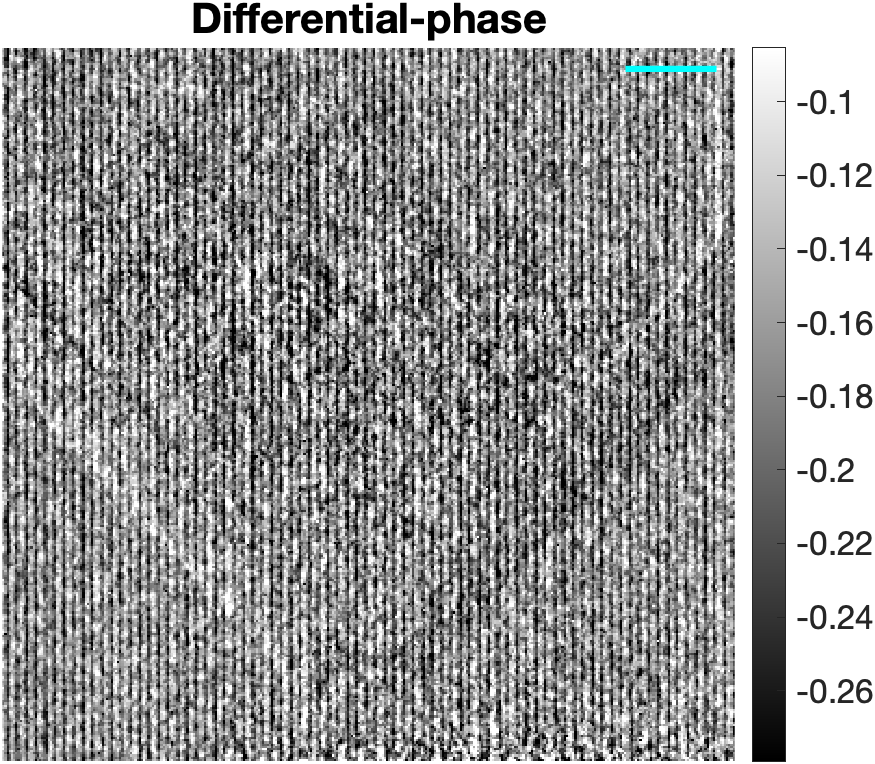}
        \subcaption{1st Harmonic, Nominal Steps, w/ Ref. Phase}
        \label{fig:PMMA/dpc_1st_only_nominal_s_with_ref_phase}
    \end{subfigure}
    \newline
    \begin{subfigure}[t]{0.49\textwidth}
        \centering
        \includegraphics[keepaspectratio=true, width=\textwidth]{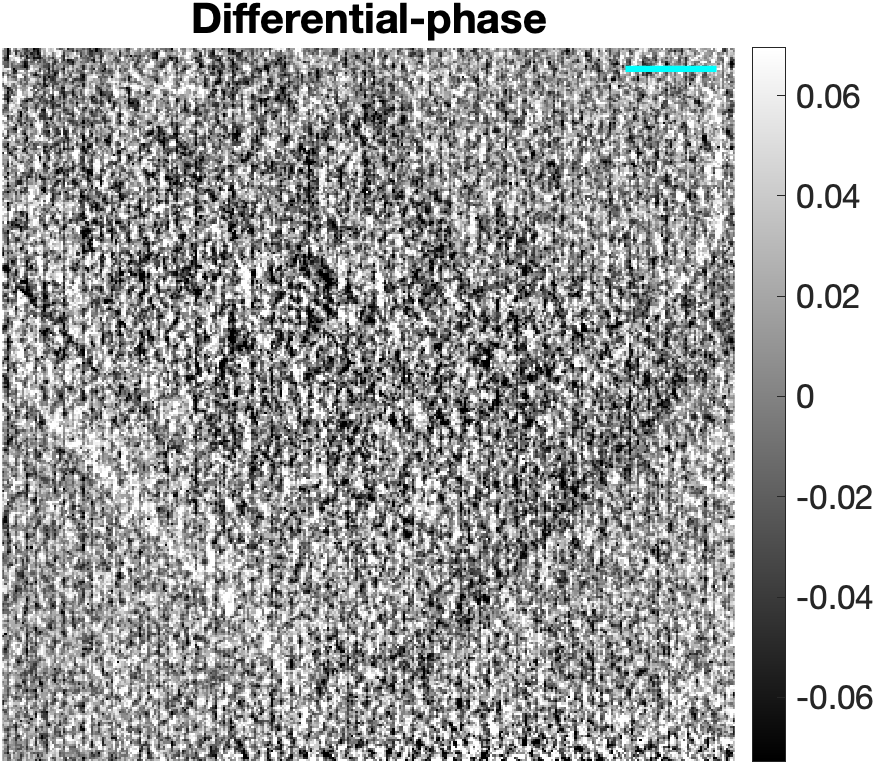}
        \subcaption{1st Harmonic, Corrected Steps, w/ Ref. Phase}
        \label{fig:PMMA/dpc_1st_only_corrected_s_with_ref_phase}
    \end{subfigure}
    \hfill
    \begin{subfigure}[t]{0.49\textwidth}
        \centering
        \includegraphics[keepaspectratio=true, width=\textwidth]{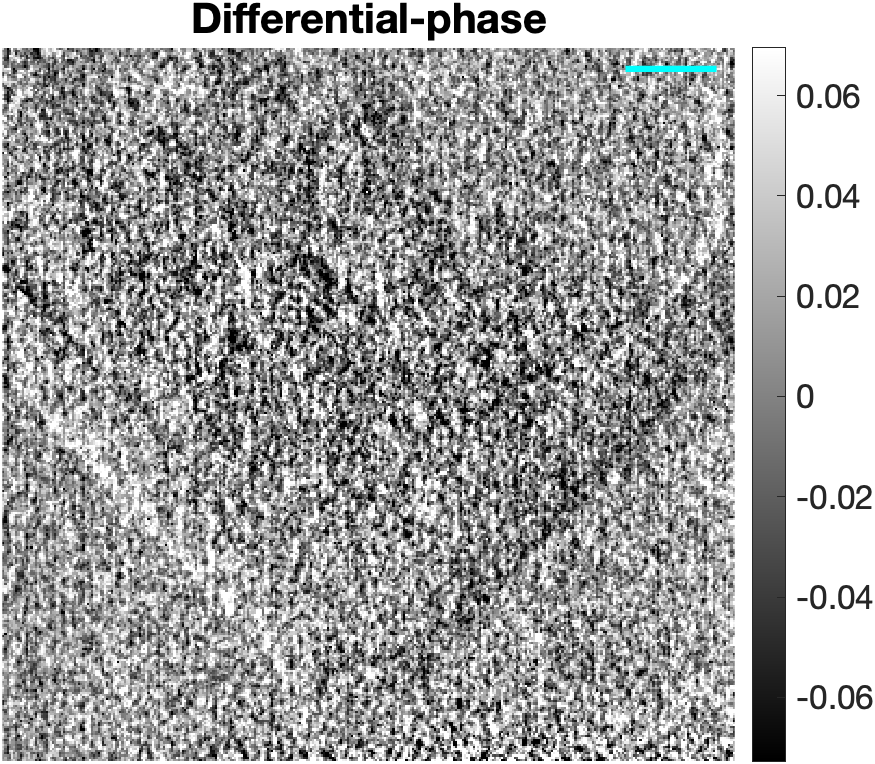}
        \subcaption{Multi-Harmonic, Corrected Steps, w/ Ref. Phase}
        \label{fig:PMMA/dpc_combined_corrected_s_with_ref_phase}
    \end{subfigure}
    \caption{Differential-phase Image calculated using 1st harmonic or multi-harmonic methods, nominal or corrected phase steps, and with or without reference phase.  A $1 \: mm$ scalebar is shown.}
    \label{fig:PMMA/dpc_comparison}
\end{figure}

In Figure \ref{fig:PMMA/dpc_comparison}, the effect of the three algorithm changes on the differential-phase image of the PMMA microspheres is shown.  Figure \ref{fig:PMMA/dpc_1st_only_nominal_s_no_ref_phase} shows the differential-phase image using the standard algorithm, with only a single harmonic, no phase wraparound correction, and nominal phase steps.  In Figure \ref{fig:PMMA/dpc_1st_only_nominal_s_with_ref_phase}, the reference phase is included in the fit of the object's phase step curve, completely removing the phase wraparound artifact (the vertical white stripes, which were $\sim 2\pi$).  In Figure \ref{fig:PMMA/dpc_1st_only_corrected_s_with_ref_phase}, the corrected phase steps are used, with a noticeable reduction in the remnant grating fringes.  Also notice the change in the colorbar, which is now centered around 0.  Before the corrected phase steps were used, the `background' differential-phase image was equal to the average difference in the corrected and nominal phase steps but was supposed to be equal to 0.  In Figure \ref{fig:PMMA/dpc_combined_corrected_s_with_ref_phase}, the multiple harmonics were used, again improving the image.

\begin{figure}
    \centering
    \begin{subfigure}[t]{0.49\textwidth}
        \centering
        \includegraphics[keepaspectratio=true, width=\textwidth]{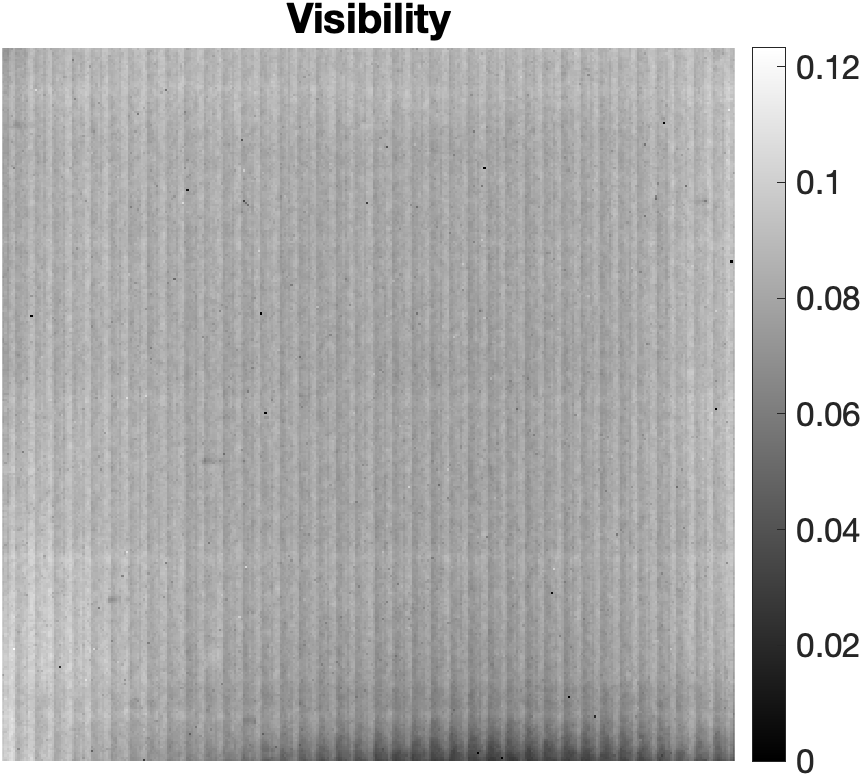}
        \subcaption{1st Harmonic, Nominal Steps}
        \label{fig:PMMA/visibility_1st_only_nominal_s}
    \end{subfigure}
    \hfill
    \begin{subfigure}[t]{0.49\textwidth}
        \centering
        \includegraphics[keepaspectratio=true, width=\textwidth]{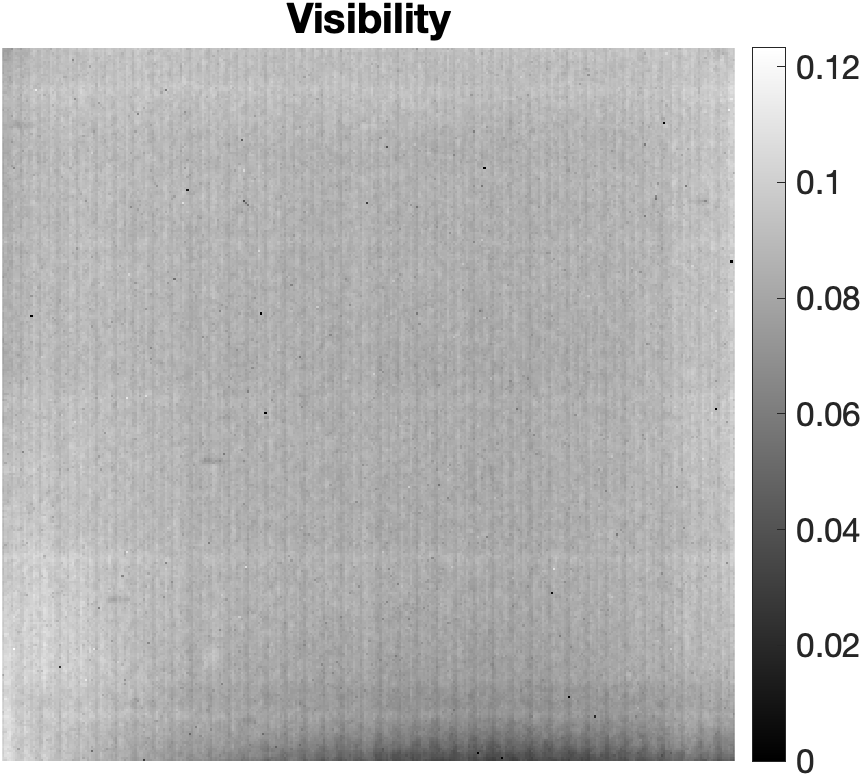}
        \subcaption{1st Harmonic, Corrected Steps}
        \label{fig:PMMA/visibility_1st_only_corrected_s}
    \end{subfigure}
    \newline
    \begin{subfigure}[t]{0.49\textwidth}
        \centering
        \includegraphics[keepaspectratio=true, width=\textwidth]{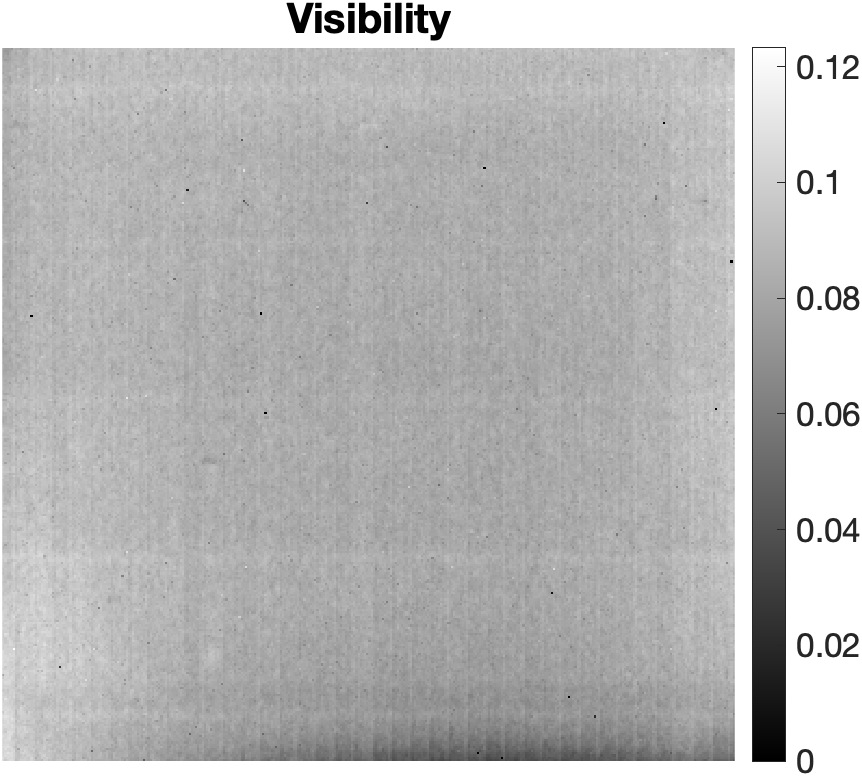}
        \subcaption{Multi-Harmonic, Nominal Steps}
        \label{fig:PMMA/visibility_combined_nominal_s}
    \end{subfigure}
    \hfill
    \begin{subfigure}[t]{0.49\textwidth}
        \centering
        \includegraphics[keepaspectratio=true, width=\textwidth]{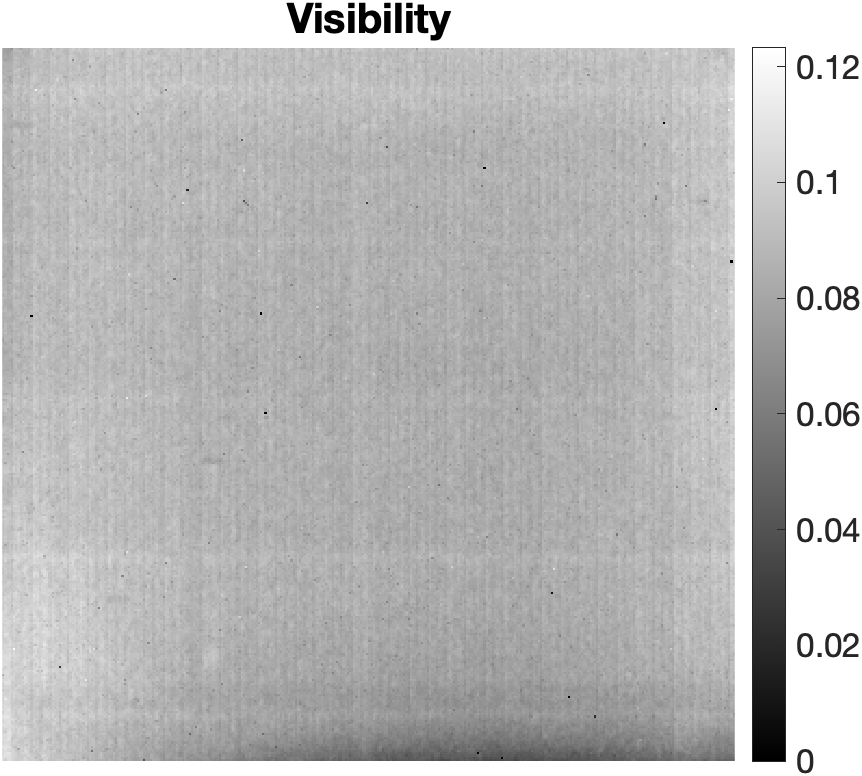}
        \subcaption{Multi-Harmonic, Corrected Steps}
        \label{fig:PMMA/visibility_combined_corrected_s}
    \end{subfigure}
    \caption{Comparison of visibility of the reference phase stepping curve for each pixel, $V^r$, calculated using nominal or corrected phase steps and using 1st harmonic or multi-harmonic methods.}
    \label{fig:PMMA/visibility_comparison}
\end{figure}

\begin{figure}
    \begin{subfigure}[t]{0.49\textwidth}
        \centering
        \includegraphics[keepaspectratio=true, width=\textwidth]{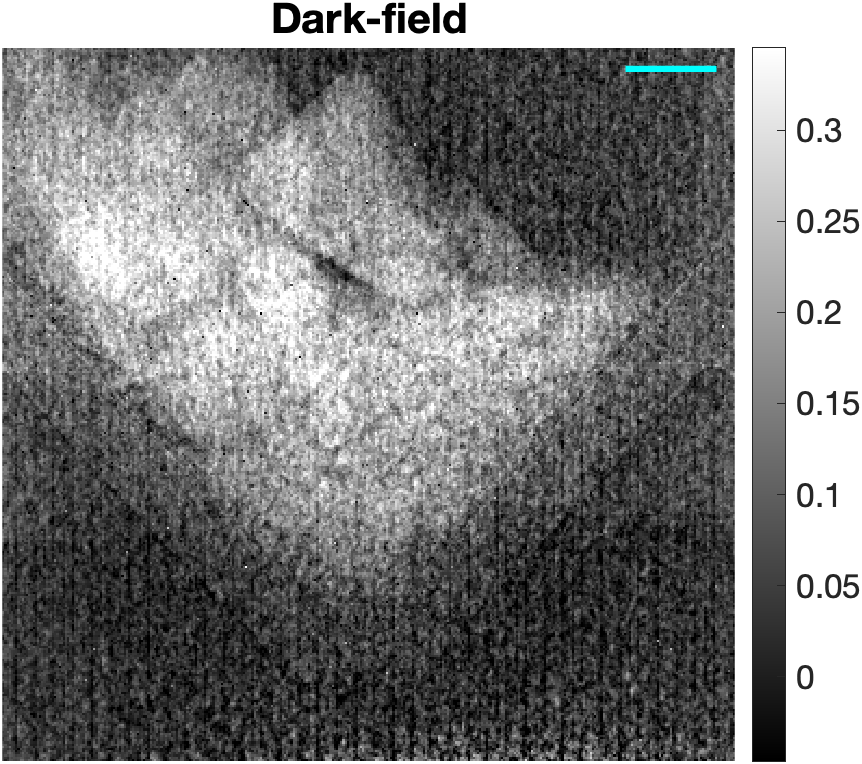}
        \subcaption{1st Harmonic, Nominal Steps}
        \label{fig:PMMA/darkfield_1st_only_nominal_s}
    \end{subfigure}
    \hfill
    \begin{subfigure}[t]{0.49\textwidth}
        \centering
        \includegraphics[keepaspectratio=true, width=\textwidth]{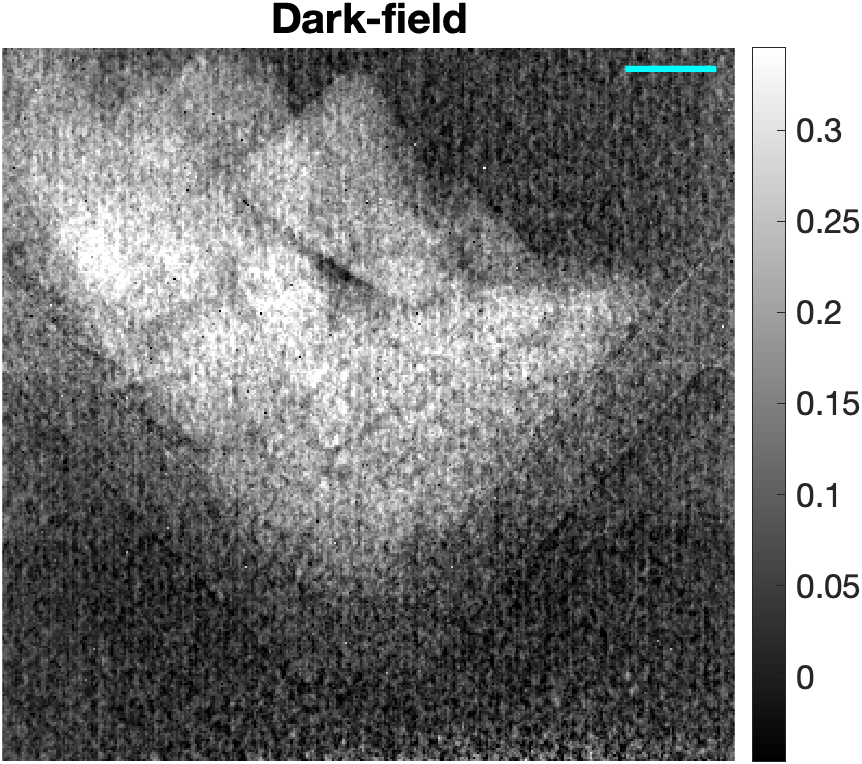}
        \subcaption{1st Harmonic, Corrected Steps}
        \label{fig:PMMA/darkfield_1st_only_corrected_s}
    \end{subfigure}
    \newline
    \begin{subfigure}[t]{0.49\textwidth}
        \centering
        \includegraphics[keepaspectratio=true, width=\textwidth]{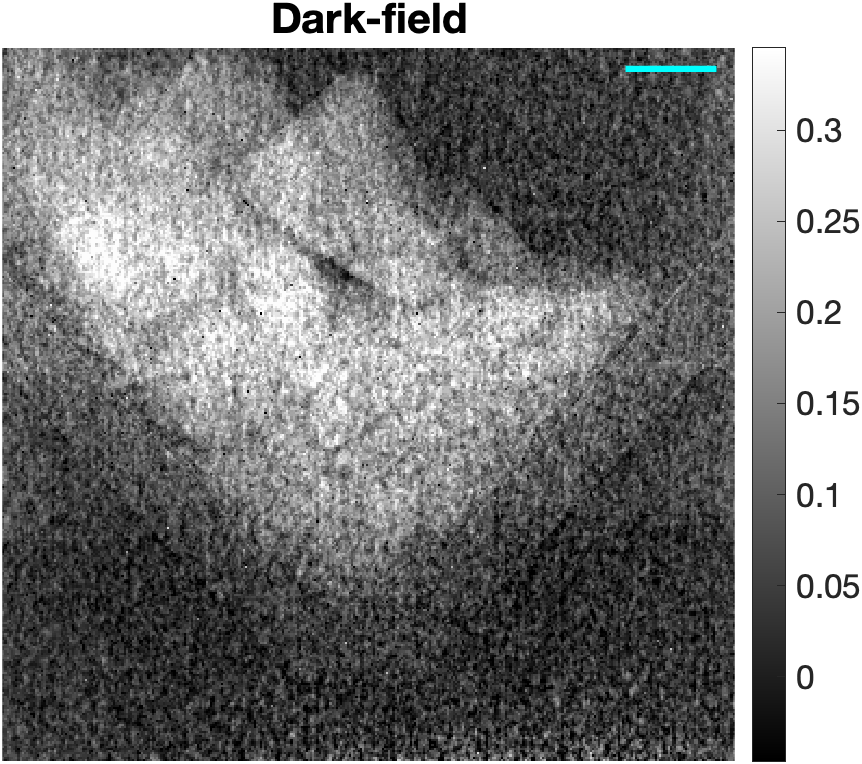}
        \subcaption{Multi-Harmonic, Nominal Steps}
        \label{fig:PMMA/darkfield_combined_nominal_s}
    \end{subfigure}
    \hfill
    \begin{subfigure}[t]{0.49\textwidth}
        \centering
        \includegraphics[keepaspectratio=true, width=\textwidth]{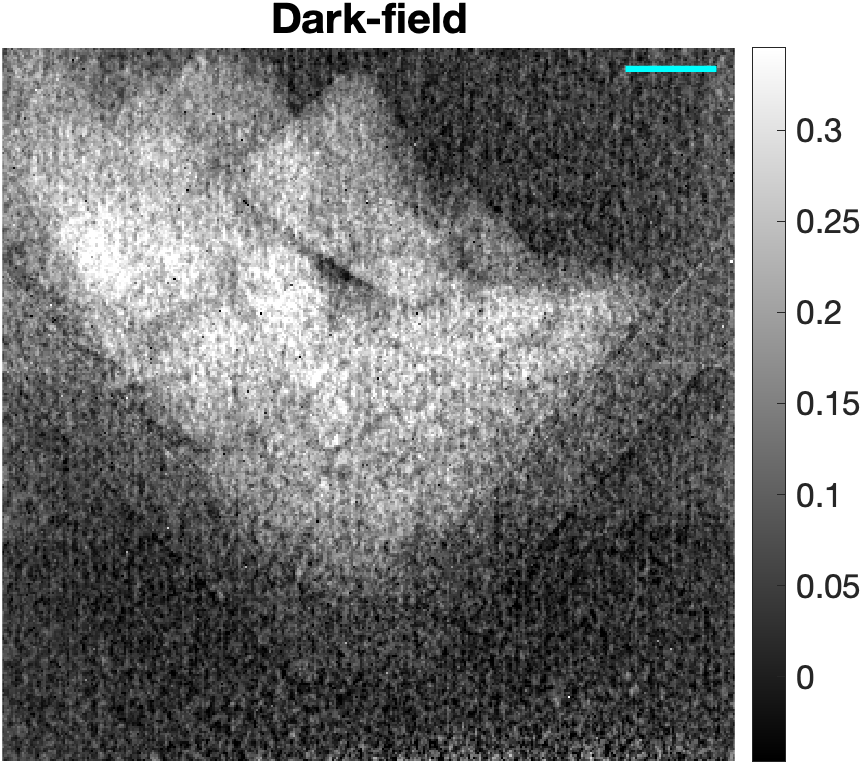}
        \subcaption{Multi-Harmonic, Corrected Steps}
        \label{fig:PMMA/darkfield_combined_corrected_s}
    \end{subfigure}
    \caption{Dark-field Image shown on a -log scale calculated using nominal or corrected phase steps and 1st harmonic or multi-harmonic methods.  A $1 \: mm$ scalebar is shown.}
    \label{fig:PMMA/darkfield_comparison}
\end{figure}

For the visibility, only the phase step corrections and multiple harmonics affect the dark-field images.  In Figure \ref{fig:PMMA/visibility_comparison}, the visibility calculated using just the 1st harmonic and nominal steps, 1st harmonic and corrected steps, multi-harmonic and nominal steps, and multi-harmonic and corrected steps are shown.  This highlights how both the corrected steps and introduction of multiple harmonics reduces the remnant fringes in the measured visibility.  In Figure \ref{fig:PMMA/darkfield_comparison}, the effect of the corrected phase steps and multiple harmonics on the dark-field image of the PMMA microspheres are shown.  In Figure \ref{fig:PMMA/darkfield_1st_only_nominal_s}, the standard algorithm is used, with a single harmonic and nominal phase steps.  In Figure \ref{fig:PMMA/darkfield_1st_only_corrected_s}, the phase steps were corrected and a single harmonic was used, and in Figure \ref{fig:PMMA/darkfield_combined_nominal_s}, multiple harmonics were used with the nominal phase steps.  In both cases, the remnant grating fringes were reduced, highlighting the importance of both changes.  In Figure \ref{fig:PMMA/darkfield_combined_corrected_s}, the phase steps were corrected and multiple harmonics were used.

\begin{figure}
    \begin{subfigure}[t]{\textwidth}
        \centering
        \includegraphics[keepaspectratio=true, width=\textwidth]{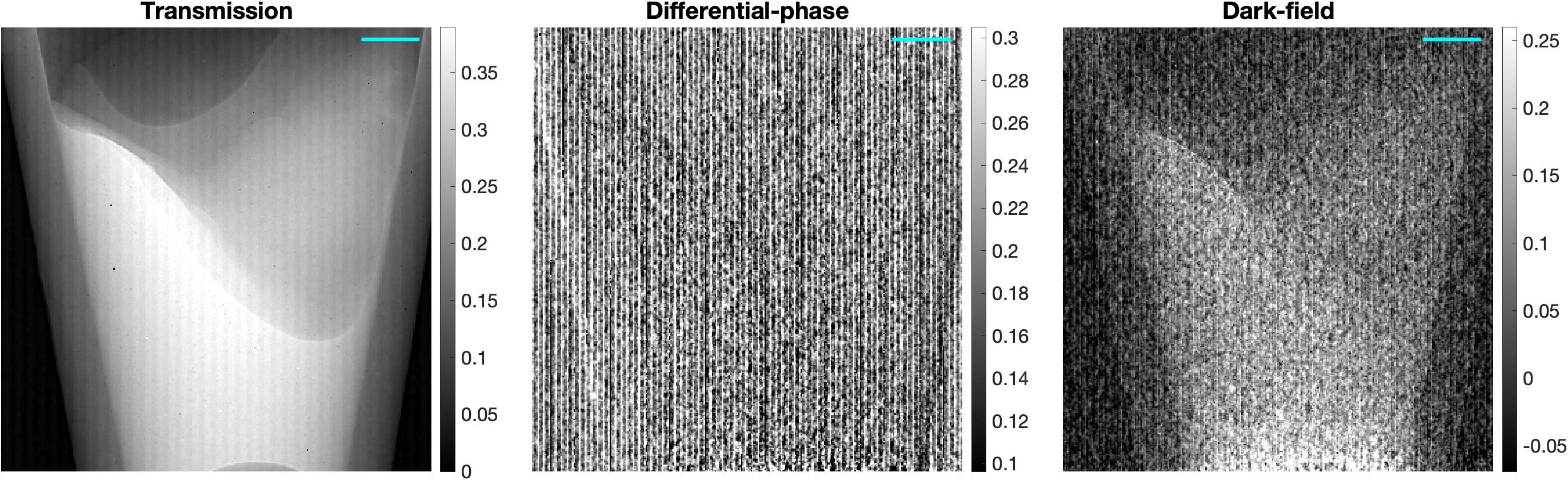}
        \subcaption{Original Method: 1st Harmonic, Nominal Steps, Without Reference Phase}
        \label{fig:Mouse_Lung/all_images_original_method}
    \end{subfigure}
    \newline
    \begin{subfigure}[t]{\textwidth}
        \centering
        \includegraphics[keepaspectratio=true, width=\textwidth]{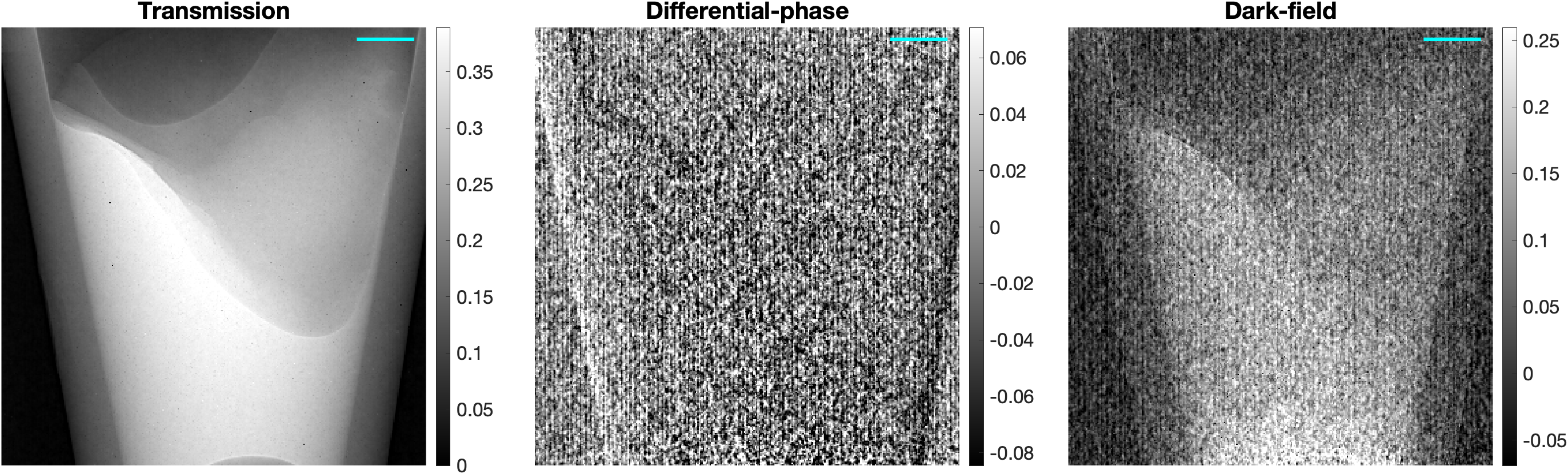}
        \subcaption{New Method: Multi-Harmonic, Corrected Steps, With Reference Phase}
        \label{fig:Mouse_Lung/all_images_new_method}
    \end{subfigure}
    \caption{Transmission, differential-phase, and dark-field comparison between the standard (top) and modified (bottom) methods for an formalin fixed ex vivo mouse lung sample.  A $1 \: mm$ scalebar is shown.}
    \label{fig:Mouse_Lung/image_comparison}
\end{figure}

\begin{figure}
    \begin{subfigure}[t]{\textwidth}
        \centering
        \includegraphics[keepaspectratio=true, width=\textwidth]{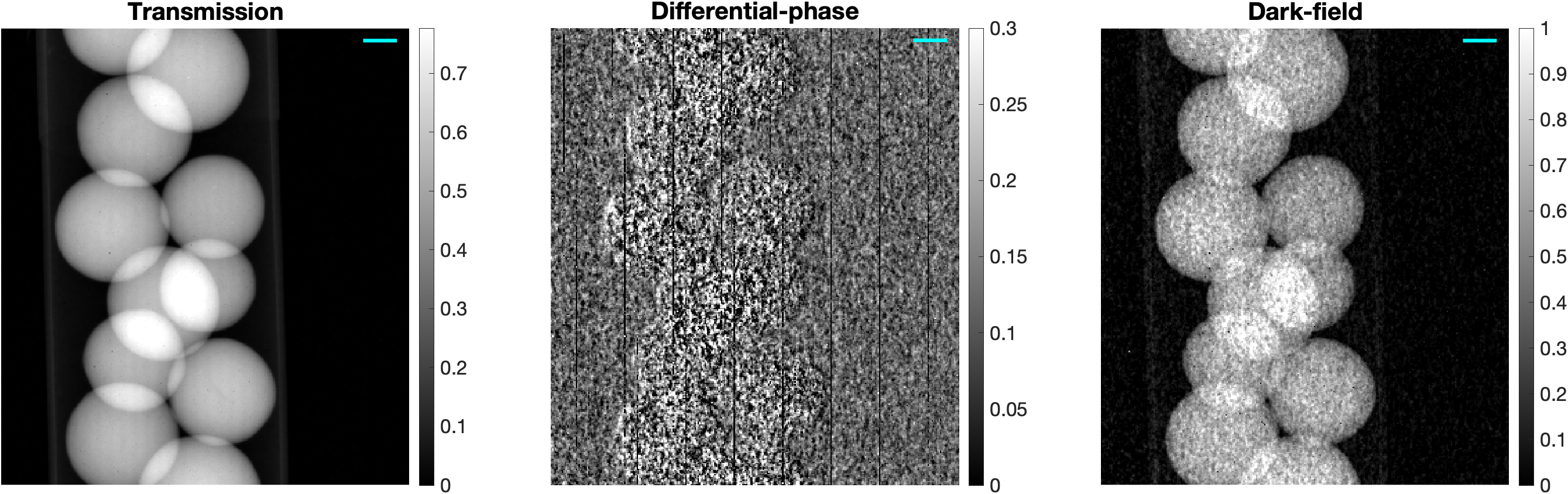}
        \subcaption{Original Method: 1st Harmonic, Nominal Steps, Without Reference Phase}
        \label{fig:SAS_90/all_images_original_method}
    \end{subfigure}
    \newline
    \begin{subfigure}[t]{\textwidth}
        \centering
        \includegraphics[keepaspectratio=true, width=\textwidth]{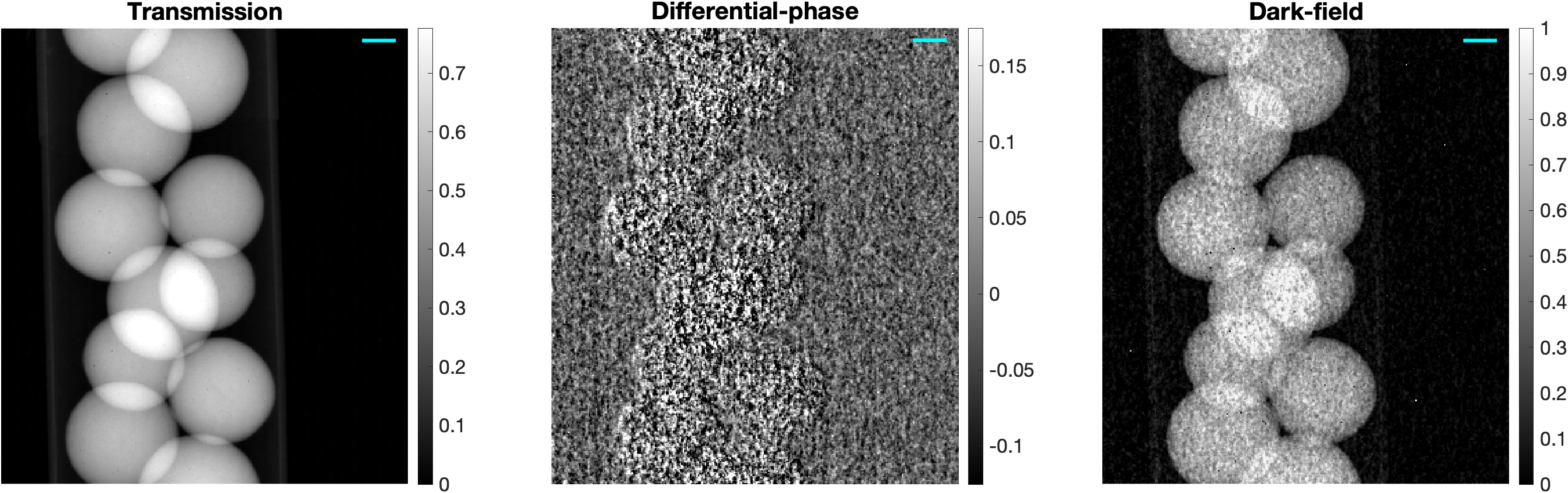}
        \subcaption{New Method: Multi-Harmonic, Corrected Steps, With Reference Phase}
        \label{fig:Mouse_Lung/all_images_new_method}
    \end{subfigure}
    \caption{Transmission, differential-phase, and dark-field comparison between the standard (top) and modified (bottom) methods for a porous alumina compound.  A $1 \: mm$ scalebar is shown.}
    \label{fig:SAS_90/image_comparison}
\end{figure}

The cumulative effect of all three algorithm changes is shown in Figures \ref{fig:Mouse_Lung/image_comparison} and \ref{fig:SAS_90/image_comparison} for the images of the ex vivo formalin fixed mouse lung and porous alumina compound.  In both cases, artifact reduction in all three images is seen.  For the alumina, the phase wraparound artifact presents itself as black stripes instead of white stripes, simply because the phase step direction was reversed for that acquisition.

\section{Discussion}
\label{sec:discussion}

We presented a non-iterative method for correcting phase step positions and included multiple harmonics in our phase stepping analysis to remove grating remnant artifacts in transmission, differential-phase, and dark-field images.  We also presented a method to remove phase wraparound artifacts in differential-phase images by including the reference phase directly into the fit of the sample's phase stepping curve.  While the methods presented were only applied to images taken with a Modulated Phase Grating Interferometer, they could easily be applied to images taken with a Talbot-Lau Interferometer or Dual Phase Grating Interferometer.

The methods presented greatly reduced the image artifacts in all three images for a variety of samples, highlighting the robustness of the image recovery algorithm.  For the phase step corrections of the sample's phase stepping curve, the entire image of each phase step was used for fitting the 2D image to a sine wave.  In limited scenarios, there is the possibility of dramatic attenuation or visibility reduction where the fit of the 2D image would not yield a valid phase, leading to incorrect phase step estimation.  In these instances, one could reduce the area of the analysis to a region of the image with high visibility fringes, but this was not necessary for our analysis.  While the previous literature has also focused on dose fluctuations, we observed less than $0.12 \%$ fluctuations of the average value for each phase step ($a_0$ from Equation \ref{eq:single_harmonic_single_image_fit}), so this was not included in this analysis.

In previous studies, we have had to reacquire the reference phase stepping curve for every sample we have imaged.  This was done to ensure the alignment of the reference and sample phase stepping curves to reduce artifacts.  With the methods presented, the reference phase stepping curve is only acquired once for each geometry, greatly reducing the required imaging time for each sample.  For future clinical systems, this may alter or reduce the requirement of quality assurance for the phase stepping motors.

With the introduction of a multi-harmonic analysis, there is the potential to isolate higher harmonic dark-field images and analyze them separately from the first harmonic image.  This analysis may prove useful for simultaneous imaging of multiple autocorrelation lengths, so long as the visibility of each harmonic is high enough.  There is also the potential to include more than two harmonics in the analysis, but overfitting may prove to be an issue.

\section{Conclusions}
\label{sec:conclusions}

X-ray interferometry has the potential for a wide variety of clinical and industrial applications, but unique image artifacts are introduced when assuming the phase stepping positions are evenly spaced and when assuming the phase stepping curves and perfectly sinusoidal.  By modifying existing algorithms, we have greatly reduced grating remnant and phase wraparound artifacts in transmission, differential-phase, and dark-field images taken with a Modulated Phase Grating Interferometer.

\section{Acknowledgments}

This work is funded in part by NIH NIBIB Trail-blazer Award 1-R21-EB029026-01A1.  The authors thank the Inhalation Research Facility of the School of Veterinary Medicine Louisiana State University, for technical support.  We also thank the National Science Foundation for the support of CBD through the REU site at Louisiana State University in the Department of Physics and Astronomy (NSF Grant No. 2150445).  We thank Dr. Kerry Dooley of the Department of Chemical Engineering, Louisiana State University for the porous alumina compounds.

\section{Author Contributions}

HCM designed and implemented the algorithms and wrote the manuscript. CBD contributed to the development and analysis of multi-harmonic corrections. VLF helped with the phase-step corrections. KH and LGB provided guidance with experiments. AN prepared and provided the mouse lung sample.  JD provided overall supervision and contributed to all aspects of this project.  All authors reviewed the manuscript.

\section{Competing Interests}

JD, KH, and LGB are inventors of two patents related to the Modulated Phase Grating Interferometry \cite{bib:MPGPatent1,bib:MPGPatent2}. JD and HCM are inventors in a provisional patent related to X-ray interferometry, filed through Louisiana State University. All of the authors confirm that their work adheres to the ethical guidelines and standards for transparency and objectivity in conducting and reporting research.

\newpage

\printbibliography

\end{document}